\documentclass[twocolumn,superscriptaddress,showpacs,prl,aps,amsmath,amssymb,nofootinbib]{revtex4-1}
\usepackage{graphicx}
\usepackage{amssymb}
\usepackage{bm}
\usepackage{color}

\newcommand{\Msol}{M_\odot}

\newcommand{\beq}{\begin{equation}} 
\newcommand{\eeq}{\end{equation}} 
\newcommand{\beqn}{\begin{eqnarray}} 
\newcommand{\eeqn}{\end{eqnarray}}

\newcommand{\zD}{{\raise1.0ex\hbox{${}^{\ \circ}$}}\!\!\!\!\!D}
\newcommand{\alone}{{\raise0.5ex\hbox{${}^{\ 1}$}}\!\!\!\!\alpha}

\newcommand{\compa}{M/R}

\newcommand{\nalam}{\mathrel{\raise0.9ex\hbox{$^\lambda$}\mkern-14mu
\lower0.0ex\hbox{$\nabla$}}}

\newcommand{\zeroD}{{\raise1.0ex\hbox{${}^{\ \circ}$}}\!\!\!\!\!D}

\newcommand{\zLap}{{\raise1.0ex\hbox{${}^{\ \circ}$}}\!\!\!\!\Delta}
\newcommand{\zna}{{\raise1.0ex\hbox{${}^{\ \circ}$}}\!\!\!\!\!\nabla}
\newcommand{\zS}{{\raise1.0ex\hbox{${}^{\ \circ}$}}\!\!\!\!\!S}

\newcommand{\cocal}{{\sc cocal}}

\newcommand{\Msphmax}{{M^{\rm SPH}_{\rm max}}}
\newcommand{\Mtrimax}{{M^{\rm TR}_{\rm max}}}
\newcommand{\Maximax}{{M^{\rm AX}_{\rm max}}}

\begin{document}

\title{
Do triaxial supramassive compact stars exist?}

\author{K\=oji Ury\=u}
%\email{uryu@sci.u-ryukyu.ac.jp}
\affiliation{Department of Physics, University of the Ryukyus, Senbaru, Nishihara, 
Okinawa 903-0213, Japan}

\author{Antonios Tsokaros}
\affiliation{Institut f\"ur Theoretische Physik, Johann Wolfgang Goethe-Universit\"at, 
Max-von-Laue-Strasse 1, 60438 Frankfurt am Main, Germany}
%\email{tsokaros@th.physik.uni-frankfurt.de}
%\email{atsok@aegean.gr}
%\affiliation{Department of I.C.S.E., University of Aegean, Karlovassi 83200, Samos, Greece} 

\author{Luca Baiotti}
\affiliation{Graduate School of Science, Osaka University, 560-0043 Toyonaka, Japan}

\author{Filippo Galeazzi}
\affiliation{Institut f\"ur Theoretische Physik, Johann Wolfgang Goethe-Universit\"at, 
Max-von-Laue-Strasse 1, 60438 Frankfurt am Main, Germany}

\author{Noriyuki Sugiyama}
\affiliation{Department of Mathematical Sciences, University of Wisconsin-Milwaukee, 
Milwaukee, Wisconsin 53211}

\author{Keisuke Taniguchi}
\affiliation{Department of Physics, University of the Ryukyus, Senbaru, Nishihara, 
Okinawa 903-0213, Japan}

\author{Shin'ichirou Yoshida}
\affiliation{Department of Earth Science and Astronomy, Graduate School of Arts and Sciences, 
The University of Tokyo, Komaba, Tokyo 153-8902, Japan}

\date{\today}

\begin{abstract}
We study quasiequilibrium solutions of triaxially deformed 
rotating compact stars -- a generalization of Jacobi 
ellipsoids under relativistic gravity and compressible equations 
of state (EOS).  For relatively stiff (piecewise) polytropic EOSs, 
we find supramassive triaxial solutions whose masses exceed 
the maximum mass of the spherical solution, but are always lower than 
those of axisymmetric equilibriums.  
The difference in the maximum masses of triaxial and axisymmetric solutions 
depends sensitively on the EOS.  If the difference turns out to be only 
about $10\%$, it will be strong evidence that the EOS of high density 
matter becomes substantially softer in the core of neutron stars.  
This finding opens a novel way to probe phase transitions of 
high density nuclear matter using detections of gravitational waves 
from new born neutron stars or magnetars under fallback accretion.  
\end{abstract}

\maketitle

\paragraph{Introduction.}
\label{sec:int}
\hskip -10pt
---\,Maclaurin spheroids and Jacobi ellipsoids, 
classical solutions of self-gravitating and uniformly 
rotating incompressible fluids in equilibrium, are 
the first two models of rapidly rotating stars.  
As the rotation of an equilibrium configuration is increased, 
a sequence of triaxial Jacobi ellipsoids branches off 
from that of axisymmetric Maclaurin spheroids where the 
ratio of kinetic to gravitational energies reach 
$T/|W|\sim 0.14$ (see, e.g.~\cite{MaclaurinBifurcation}).  
This led to historical mathematical studies including 
Poincar\'e's bifurcation theory \cite{PoincareLamb}.  
A generalization of the Maclaurin spheroids, 
and other stationary axisymmetric equilibriums, to 
the case of relativistic gravity have been fully 
investigated in \cite{Meinel_book}.  
From a point of view of relativistic astrophysics, 
it is also important to include compressibility of 
the fluid; realistic neutron stars are modeled 
as axisymmetric and uniformly rotating configurations 
associated with equations of state (EOSs) of 
high density nuclear matter 
(see e.g.~\cite{2013rrs..book.....F,2013gere.book.....S}).

It is not so surprising that a relativistic 
generalization of Jacobi ellipsoids, even for the case with 
compressible fluid, has been of little astrophysical 
interest, because of the following four difficulties.  
Firstly, such non-axisymmetric, triaxially deformed, 
solutions can not be stationary equilibriums 
due to the back reaction of gravitational waves 
\cite{Chandrasekhar:1992pr,ellipsoidal}.\footnote{
Hereafter we use a term 'triaxially deformed' or 
simply 'triaxial' star rather than 'ellipsoid', since 
the configurations are no longer an exact ellipsoid 
in relativistic gravity or for compressible fluids.  
The triaxial configurations in this paper possess 
the tri-planar symmetry with respect to three orthogonal 
x, y, and z planes.}  
Secondly, there should be a highly efficient mechanism 
to spin up the compact star as fast as $T/|W|\sim 0.14$.  
Thirdly, in realistic high density nuclear matter, the viscosity 
may not be strong enough to bring a flow field to 
uniform rotation within a time scale shorter than 
the time scale of gravitational radiation \cite{Chandrasekhar:1992pr,ellipsoidal}.  
Fourthly, even in Newtonian gravity, such a triaxial sequence 
does not exist for the gaseous EOSs unless the EOS is stiff 
enough.  For the case with polytropic EOS, $p=K\rho^\Gamma$, 
the triaxial sequence only exists in the range $\Gamma \agt 2.24$, 
where $p$ is the gas pressure, $\rho$ the 
(rest mass) density, $K$ the adiabatic constant, 
and $\Gamma$ the adiabatic index \cite{1964ApJ...140..552J}.  
Even for such stiff EOS, say $2.24 \alt \Gamma \alt 4$, 
the triaxial sequence is terminated at the mass shedding limit 
not very far away from the branching point as its angular 
momentum is increased \cite{Jacobiseq}.  

Although a couple of Kuiper Belt objects are likely to rotate 
rapidly enough to become Jacobi ellipsoids \cite{KBO}, 
it is still inconclusive whether such 
triaxially deformed rapidly rotating configuration is 
realized or not for compact objects such as neutron stars.  
However, the last two difficulties above may be avoided.  
There are various types of phenomenologically 
derived high density nuclear matter EOSs, some of which 
may be approximated fairly accurately by polytropic 
or better by piecewise polytropic EOSs with 
$\Gamma$ as large as $\Gamma\sim 3-4$ \cite{Read:2008iy}.  
Viscosity of neutron star matter, which is normally expected 
to be weak, may be enhanced by magnetic 
effects and/or high temperature \cite{NSviscosity}.

Moreover, in a recent paper \cite{2011ApJ...736..108P}, 
Piro and Ott 
have shown that the supernova fallback accretion 
may spin up a newly formed neutron star 
associated with the strong magnetic field $B \alt 5\times10^{14}$\,G 
as fast as the above criteria $T/|W|\sim0.14$ 
for $\sim 50$ -- $200$\,s until the star collapses to a black hole.  
Therefore, there is a possibility that such triaxially 
deformed compact stars may be formed transiently after massive stellar core 
collapses.  Once such triaxial star is formed, the amount of 
gravitational wave emission is enormous, from which 
we could extract properties of high density nuclear matter.  
Piro and Thrane have estimated the detectability of 
gravitational waves from triaxially deformed compact stars 
within the fallback accretion senario for the case of the advanced 
LIGO detector \cite{Harry:2010zz} as $\sim 17$ Mpc using a realistic 
excess cross-power search algorighm \cite{2012ApJ...761...63P}.\footnote{
The amplitude of gravitational waves from triaxial stars 
is typically \cite{Lai:1994ke}, 
\[
h \sim 9.1\times 10^{-21}\left(\frac{30\,\mbox{Mpc}}{D}\right)
\left(\frac{M}{1.4\Msol}\right)^{3/4}
\left(\frac{R}{10\,\mbox{km}}\right)^{1/4}
f^{-1/5}, 
\]
where $D$, $M$, $R$, and $f$ are, respectively, the distance to the source, 
the source mass, the mean radius, and the wave frequency in Hz.
}
This scenario motivates us to further investigate the properties of 
triaxially deformed compact stars.\footnote{A magnetic field 
$B \alt 5\times10^{14}$\,G is not strong enough to alter 
the hydrostatic equilibrium of rotating compact stars.}

In our previous calculations \cite{Huang:2008vp,Uryu:2016dqr}, 
it was apparent that the triaxial sequence becomes shorter 
(that is, a smaller deformation is allowed) 
for the case with higher compactness.  
For certain EOSs, it is even unclear whether there exist 
supramassive triaxial solutions whose masses are higher than 
the maximum mass of the spherically symmetric solutions of 
Tolman-Oppenheimer-Volkov (TOV) equations, just like 
the case for axisymmetric uniformly rotating solutions.  
In this letter, we present for the first time a systematic 
study of the classical problem for computing triaxially deformed 
uniformly rotating stars in general relativistic gravity, 
and elucidate the properties of the quasiequilibrium sequences 
of such rotating stars for (piecewise) polytropic EOSs 
up to their maximum mass.  

%%% 

\paragraph{A method for computing sequences of solutions.}
\label{sec:method}
\hskip -10pt
---\,We focus on computing rotating compact stars for three EOSs.  
Two of them are polytropic EOSs $p=K\rho^\Gamma$ with adiabatic 
constant $\Gamma=3$ or $4$, and the other is a two segments piecewise 
polytropic EOS $p=K_i\rho^{\Gamma_i}$ $(i=1,2)$, with 
$\Gamma_1=4$ for $\rho \leq 2 \rho_{\rm nuc}$ and 
$\Gamma_2=2.5$ for $\rho > 2 \rho_{\rm nuc}$.  
We set the interface value of the rest mass density $\rho_{\rm nuc}$ 
to be the nuclear saturation density 
$\rho_{\rm nuc} = 2.8\times 10^{14} {\rm g/cm}^3$ in cgs unit.  
We choose the value of the adiabatic constant $K$ and $K_i$ so that 
the value of the rest mass $M_0$ becomes $M_0=1.5 \Msol$ at the compactness 
$M/R=0.2$ for the TOV solution.
Physical quantities of spherically symmetric solutions at the 
maximum mass of these EOSs are presented in Table \ref{tab:TOV_solutions}.  
\footnote{
The adiabatic speed of sound $c_s:=\sqrt{dp/d\epsilon}$ 
for the polytropic EOSs with $\Gamma=3$ and $4$ 
exceeds the speed of light when the rest mass density 
$\rho/\rho_c \agt 0.898$ and $0.656$, where 
values of $\rho_c$ are tabulated in Table \ref{tab:TOV_solutions} for 
 $\Gamma=3$ and $4$, respectively.  
The results from these acausal EOSs for $\Gamma=3$ and $4$ are shown 
for a comparison with 
our piecewise polytropic EOS model with $(\Gamma_1,\Gamma_2)=(4,2.5)$ 
%%%These acausal EOSs are chosen for a comparison with 
%%%our piecewise polytropic EOS model $(\Gamma_1,\Gamma_2)=(4,2.5)$ 
which is always causal in the range of $\rho$ calculated in this letter.  } 
%%% including $\rho=\rho_c$ in the same Table \ref{tab:TOV_solutions}.}

The most accurate rotating triaxial equilibriums of compact stars 
would be computed as helically symmetric solutions associated with 
standing gravitational waves.  One can, however, truncate the 
gravitational-wave content because its contribution to the source's 
equilibrium is small, and instead compute quasiequilibrium initial data 
on a three dimensional hypersurface.  We have developed a code for 
computing such data as a part of our Compact Object CALculator \cocal\ 
code \cite{Uryu:2016dqr,cocal_BBH}.  To reduce computing 
time, we use the Isenberg-Wilson-Mathews formulation in this paper.  
Further details on the numerical method, 
as well as the definitions on physical quantities, 
are found in \cite{Uryu:2016dqr}.  
%%% and differences between the present result 
%%% and those of more accurate quasiequilibrium data 
%%% computed from the waveless formulation 
%%% are reported in \cite{Uryu:2016dqr}.  

For each EOS and for both axisymmetric and triaxial configurations, 
we compute sequences of solutions varying 
two parameters which determine the compactness (or the mass) 
and the degree of rotation.  
In practice, for the former, we choose the central 
density $\rho_c$, and for the latter, 
the axis ratio (deformation) $R_z/R_x$ for the axisymmetric 
solutions, and $R_y/R_x$ for the triaxial solutions, where 
$R_x$, $R_y$, $R_z$ are the radii along the semi-major axis.  
The $z$-axis corresponds to the axis of rotation, and 
the $x$-axis is along the longest semi-principal axis for the case of triaxial solutions.  
%
%%% Sequences of numerical solutions are calculated in the following 
%%% order.  We start iterations using a low mass spherically symmetric 
%%% solution as an initial guess.  A sequence of solutions with an increasing 
%%% deformation (decreasing the axis ratio $R_z/R_x$) 
%%% is calculated under the central density being 
%%% kept constant.  
%
%%% For the axisymmetric solutions, 
%%% the range of deformation is typically 
%%% from $R_z/R_x= 120/128$ to $72/128$, 
%%% decreasing every $8/128$ or $4/128$ steps, 
%%% and for the triaxial solutions 
%%% the equatorial axis ratio $R_y/R_x$ is 
%%% decreased from $R_y/R_x=124/128$ every $4/128$ steps 
%%% until the deformation sequence reach to the mass-shedding limit.  
%
For each deformation model, a sequence of solutions 
is calculated with increasing $\rho_c$, 
typically from $\rho_c=3.0 \times 10^{14} {\rm g/cm}^3$ to 
$3.0\times10^{15} {\rm g/cm}^3$.  
As $\rho_c$ is increased, these sequences with the fixed deformation 
may or may not be terminated at the mass-shedding limit before 
$\rho_c$ reaches $3.0\times10^{15} {\rm g/cm}^3$.  
As far as our selected EOS models are concerned, the triaxial sequences 
with fixed $R_y/R_x$ and increasing $\rho_c$ 
are always terminated at the mass shedding limit, while 
the axisymmetric sequences are terminated at the mass shedding 
limit only for the smaller $R_z/R_x$ (larger deformation) cases.

\begin{table}
\begin{tabular}{cccccc}
\hline
$\Gamma$ & $(p/\rho)_c$ & $\rho_c$ & $M_0$ & $M$ & $M/R$  \\
\hline
$3$      & $0.827497$ & $0.00415972$ & $2.24295$ & $1.84989$ & $0.316115$  \\
$4$      & $1.330409$ & $0.00322082$ & $2.88207$ & $2.24967$ & $0.355062$  \\
$(4, 2.5)$ & $0.568330$ & $0.00454117$ & $1.96013$ & $1.65738$ & $0.287213$  \\
\hline
\end{tabular}
\caption{Quantities at the maximum mass of spherically symmetric 
solutions are listed for the polytropic EOSs $p=K\rho^\Gamma$ 
with $\Gamma=3$ and $4$, and for the two segments piecewise 
polytropic EOS $p=K_i\rho^{\Gamma_i}$ with $(\Gamma_1,\Gamma_2)=(4,2.5)$.  
The adiabatic constant $K$ and $K_i$ are chosen so that the value of the rest mass 
$M_0$ becomes $M_0=1.5$ at the compactness $M/R=0.2$.\footnote{
For the relativistic (piecewise) polytropes, physical dimensions enter only 
through the constant $K$.  Dimensionless values of mass and radius are obtained 
from dividing each by a factor $K^{1/2(\Gamma-1)}$ in $G=c=1$ unit.  
}
Note that the last EOS $(\Gamma_1,\Gamma_2)=(4,2.5)$ is softer than the others.  
Values are in $G=c=M_\odot=1$ unit, and are approximated using 
2nd order interpolation of nearby 3 solutions.  To convert the units of 
the central density $\rho_c$ to cgs, multiply by 
$\Msol(G\Msol/c^2)^{-3} \approx 6.176393\times 10^{17} {\rm g}\ {\rm cm}^{-3}$.
}  
\label{tab:TOV_solutions}
\end{table}
\begin{figure*}
\begin{center}
\includegraphics[height=44mm]{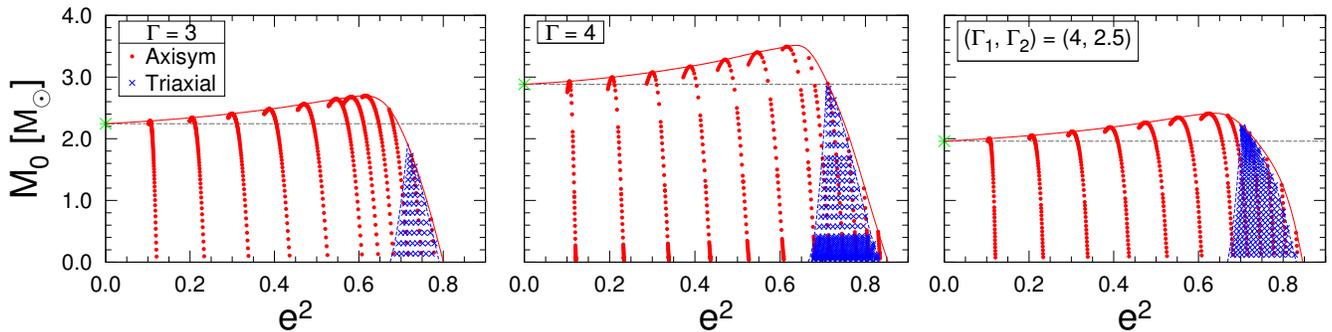}
\caption{
The rest mass $M_0$ is plotted against the square of the eccentricity 
(in proper length) $e^2:=1-(\bar{R}_z/\bar{R}_x)^2$ 
for axisymmetric (red dots) and triaxial (blue crosses) solutions 
of uniformly rotating compact stars.  Solid (red) and 
dashed (blue) envelope curves are polynomial fits to the extrapolated 
limiting solutions.  Left to right panels correspond to the results of 
polytropic EOSs $\Gamma=3$ and $4$, and piecewise polytropic EOS 
$(\Gamma_1,\Gamma_2)=(4,2.5)$, respectively.  Solutions above the 
horizontal dashed line in each panel are supramassive, $M_0 > \Msphmax$.  
In each panel, the left fitted curve to the triaxial solutions (blue dashed) 
corresponds to the bifurcation points, and the right to the mass shedding 
(Roche) limits.}  
\label{fig:sequence}
\end{center}
\end{figure*}
\begin{figure}
\begin{center}
\includegraphics[height=50mm]{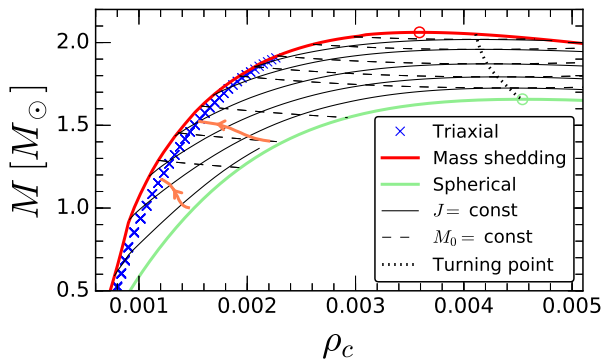}
\caption{
The gravitational (Arnowitt-Deser-Misner) mass $M$ is plotted against 
the central density $\rho_c$ for the same model with the right panel of 
Fig.~\ref{fig:sequence}, $(\Gamma_1,\Gamma_2)=(4,2.5)$.  
Solid and dashed curves are 
axisymmetric sequences with $J=\mbox{constant}$, 
and $M_0=\mbox{constant}$, respectively.  Turning points of these 
curves are indicated by a dotted curve. Top (red) and bottom (green) 
circles are the maximum of the gravitational mass for the axisymmetric 
and spherical stars, respectively.
For a reference, we draw evolutionary tracks (thick curves with arrows) 
of newly born magnetars under the fallback accretion modeled by 
Piro and Ott Eqs.\,(14)-(16) in \cite{2011ApJ...736..108P}. 
For simplicity, a spin equilibrium (Eq.\,(17) 
in \cite{2011ApJ...736..108P}) is always assumed.
}
\label{fig:M_rhoc}
\end{center}
\end{figure}

\paragraph{Results.}
\label{sec:method}
\hskip -10pt
---\,In Fig.~\ref{fig:sequence}, the rest mass $M_0$ is plotted 
with respect to the square of eccentricity in proper length 
$e^2:=1-(\bar{R}_z/\bar{R}_x)^2$ 
for the above three EOS models.  
%%%for each EOS with adiabatic 
%%%indices $\Gamma=3$, $\Gamma=4$, and $(\Gamma_1,\Gamma_2)=(4,2.5)$ from 
%%%left to right panels, respectively.  
A cross point on the vertical axis at $e^2=0$ in each panel indicates 
the maximum rest mass of a spherically symmetric solution for 
each EOS model tabulated in Table \ref{tab:TOV_solutions}.  
We notice that, for the case with $\Gamma=3$, 
the maximum mass of triaxial solutions never exceeds 
that of the spherical solutions.  For the $\Gamma=4$ case, 
the mass of all computed triaxial solutions again does not 
exceed that of the spherical solutions.  However, if we extrapolate 
the triaxial solutions closer towards the axisymmetric 
solution (a peak of dashed curves), we could find triaxial solutions 
with mass higher than the maximum mass of the spherical solutions.  
Therefore, we may conclude that supramassive triaxial 
solutions exist for $\Gamma\agt 4$, although the excess of 
mass is much lower than that of axisymmetric 
supramassive solutions.

It is important to notice that 
the deformation sequence of triaxial solutions with a constant 
rest mass $M_0$ becomes shorter for more massive (higher compactness) 
models, and hence the maximum mass of the triaxial solutions can be 
found in the vicinity of the bifurcation point of the axisymmetric 
and triaxial sequences.  This is due to the fact that 
the density distribution becomes more centrally condensed 
as the relativistic star becomes more compact, and 
hence the mass shedding limit of $M_0={\rm constant}$ sequence 
(where the matter at the equator (or at the largest radius 
for the case with the triaxial solution) brakes up) appears at  
a smaller deformation ($R_y/R_x$ closer to unity).\footnote{This is 
analogous to Newtonian rotating stars; for softer and more 
centrally condensed EOSs, rotating equilibriums reach 
the brake up velocity (Roche limit) with a smaller deformation.}  
Hereafter, we denote the maximum rest mass of the spherical 
solutions, the rotating axisymmetric sequences, and the rotating 
triaxial sequences by $\Msphmax$, 
$\Maximax$, and $\Mtrimax$, respectively.  
%%% Note that qualitative 
%%% behaviors for the ADM and Komar masses are the same as the rest mass.  

For the piecewise polytrope model with $(\Gamma_1,\Gamma_2)=(4,2.5)$ 
one might expect that, since the value of $\Gamma_2$ of this EOS 
is substantially lower than the $\Gamma=4$ polytrope, $\Mtrimax$ 
for this EOS may become lower than  $\Msphmax$ 
as in the case with $\Gamma=3$.  The right panel of 
Fig.~\ref{fig:sequence} shows that it is not the case; 
the supramassive triaxial solutions for the $(\Gamma_1,\Gamma_2)=(4,2.5)$ EOS 
do clearly exist.  
For axisymmetric solutions, $\Maximax$ exceeds $\Msphmax$ 
for each EOS around $20\%$: 
for the computed solutions in Fig.~\ref{fig:sequence}, 
the excesses are $20.2\%$, $21.2\%$, and $22.8\%$ for 
$\Gamma=3$, $\Gamma=4$, and $(\Gamma_1,\Gamma_2)=(4,2.5)$, respectively.  
On the other hand, the calculated $\Mtrimax$ 
in Fig.~\ref{fig:sequence}, which appears close to 
the bifurcation point, falls behind $\Msphmax$ 
by $-23.2\%$ and $-2.41\%$ for $\Gamma=3$ and $\Gamma=4$, 
respectively, while it exceeds $11.5\%$ for the $(\Gamma_1,\Gamma_2)=(4,2.5)$ case.  

This striking difference in the behavior between 
$\Maximax$ and $\Mtrimax$
can be understood qualitatively as follows.  
In Fig.~\ref{fig:sequence}, each consecutive point from smaller to larger 
$M_0$ in $(e^2,M_0)$ plane corresponds to a sequence $M_0(\rho_c)$ with 
a fixed axis ratio (and varying $\rho_c$).  For the axisymmetric solutions 
with $e^2 \alt 0.7$, each sequence has a turning point where $M_0$ 
reaches the maximum and then decrease as $\rho_c$ increases.  
This is related to a change of stability associated with 
the fundamental (F) mode \cite{2013rrs..book.....F}.  
As shown in Fig.~\ref{fig:M_rhoc} for the case with 
$(\Gamma_1,\Gamma_2)=(4,2.5)$, 
simultaneous turning points appear on $M(\rho_c)$ curves 
with constant $M_0$ and $J$, where $M$ is the gravitational 
(Arnowitt-Deser-Misner) mass.  
There, the axisymmetric configurations become 
radially unstable, just like a stability change at the maximum mass of 
spherical stars $\Msphmax$ \cite{Friedman:1988er,CST}.  
%
%%% \footnote{
%%% The turning points on the sequences with fixed axis ratio 
%%% are not the points where the stability changes.  
\footnote{
The criteria is known to be a sufficient condition for 
stability, and recent simulations suggest that the vanishing 
F-mode appears somewhat smaller $\rho_c$ than that determined 
by the turning point method \cite{Takami:2011zc,2013rrs..book.....F}.  
Because the point where the stability changes is placed beyond 
the maximum mass $\Maximax$, the difference between $\Maximax$ and 
the mass at the radial stability limit does not affect our discussion.  
}

The fact that the sequences $e^2 \agt 0.7$ in Fig.\ref{fig:sequence} 
are without 
turning points suggests that these solutions are not subject to 
radial instability, and the sequences 
$M_0(\rho_c)$ with fixed axis ratio terminate at the mass shedding 
(Roche) limit.  
The maximum mass of the (supramassive) axisymmetric uniformly rotating 
solutions $\Maximax$ \emph{in our definition} is expected to appear 
near this turning point and the Roche limit merge.  
$\Maximax$ 
is, hence, associated with the radial instability, and this 
excess of $\sim 20\%$ from $\Msphmax$
turned out to be almost independent 
on the EOS for the uniformly rotating case.  
In contrast, $\Mtrimax$ 
is not related to the radial instability limit 
but to the Roche limit in the range of the adiabatic 
index we are interested in.  
The Roche limit is sensitive to the stiffness of the EOS: 
the stiffer EOSs in the lower density region, say $\rho < 2\rho_{\rm nuc}$, 
prevent mass shedding to occur.  
%% near the Kepler rotation limit.  

Therefore, if, as in certain models of realistic neutron star EOSs, 
the EOS is softer for higher densities (inner core), and 
is stiffer for lower densities (outer core), 
supramassive triaxial neutron stars are formed.

\begin{figure}
\begin{center}
\includegraphics[height=27mm]{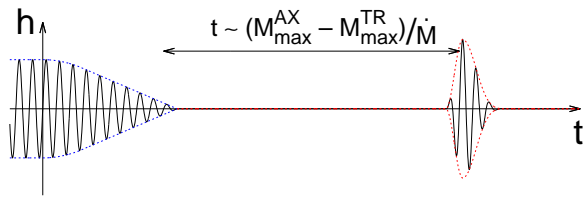}
\caption{An illustrative gravitational waveform 
from a compact star under fallback accretion.  
A diminishing periodic wave from a triaxially deformed 
rotating star is followed by a burst wave from a collapse.}
%
%%%A periodic wave from a triaxially deformed 
%%%rotating star is diminished and followed by 
%%%a burst wave from a collapse.}
\label{fig:waveform}
\end{center}
\end{figure}

\paragraph{Discussion.}
\label{sec:discussion}
\hskip -10pt
---\,In \cite{2011ApJ...736..108P}, it is demonstrated that the accretion rate of 
fall back material to a new born strongly magnetized neutron star 
in the supernova remnant can be as high as 
$\dot{M} \sim 10^{-4}$-$10^{-2} \Msol/{\rm s}$ and 
transport enough angular momentum to spin up the neutron star and to cause 
the onset of a non-axisymmetric instability 
(see, Fig.\ref{fig:M_rhoc}).  
Assuming the accretion rate to be constant at this rate, 
we expect that the star stays at the temperature, $T$, of the order of 
$T\sim 10^9$K, because of the continuous emission of thermal neutrinos 
\cite{2000MNRAS.316..917Y}.  At such high temperature, 
the bulk viscosity of neutron star matter dominates over the shear viscosity 
as their temperature dependences are $\propto T^6$ and $\propto T^{-2}$ for 
the bulk and shear, respectively.  The bulk viscosity also dominates 
over the gravitational waves so that it drives the star to a non-axisymmetric 
figure -- establishing a Jacobi-like configuration \cite{NSviscosity}.  

Detectability of gravitational waves from such accreting neutron stars 
(or magnetars) has been discussed in \cite{2012ApJ...761...63P}, in which the gravitational 
waveform is modeled as periodic waves from Jacobi ellipsoids with 
increasing mass.  This scenario is modified for the case with 
compressible EOS.  Our finding suggests that the periodic 
gravitational wave signal from triaxially deformed neutron stars 
would be terminated at the time when the mass approaches 
$\Mtrimax$.  It is likely that the accretion continues 
with the same rate after the disappearance of the periodic 
signal as the mass increases beyond $\Mtrimax$.  
Then, within $10$-$1000$s, we expect a gravitational 
wave burst, or a prompt emission of some electromagnetic 
signal from the collapse of the axisymmetric neutron star 
to form a black hole as the mass grows over $\Maximax$.  

Modeling of the waveform from such an accreting triaxial compact star, 
which may look like the one in Fig.~\ref{fig:waveform}, 
is beyond our scope in this letter.  
However, we stress the qualitative importance of the detection of 
such gravitational waves and its implication for the 
EOS of the high temperature side of the high density neutron star matter.  
From the data analysis of periodic gravitational waves emitted from 
the accreting triaxially deformed neutron stars, we could determine 
the maximum mass of the triaxial solution $\Mtrimax$ (which may or 
may not be supramassive) and the mass accretion rate $\dot{M}$.  
The maximum mass of the axisymmetric supramassive solution $\Maximax$ 
may also be determined from the duration between the disappearance 
and the burst of gravitational wave signals (or from the burst waveform itself).  
The time until a collapse to form a BH may be detected also through 
other electromagnetic signals.  
And most importantly, the gap between these two signals carries clear 
information on the EOS of high density neutron star matter.  
If the value of $\Maximax$ turned out to be only about 
$10\%$ larger than $\Mtrimax$, it would be strong evidence 
for the fact that the EOS of high density neutron star matter 
is substantially softer in the core of neutron stars.  

It should be noted that our (piecewise) polytropic EOS is understood 
as a parametrization of various types of nuclear EOS.  
A variety of microphysics of high density nuclear matter can be 
integrated into the adiabatic indices $\Gamma_i$, 
the dividing densities at the interfaces of successive segments, 
and an adiabatic constant of one of segments.  
As demonstrated in \cite{Read:2008iy}, 
it is a best practice to introduce such piecewise polytropic EOS 
with a minimal number of segments to parametrize realistic EOSs to 
constrain them through gravitational wave observations.  
However, according to \cite{Read:2008iy} for the case with binary neutron 
star inspirals, one EOS parameter may be constrained from the gravitational 
wave observations by advanced LIGO detectors, and two by Einstein Telescope 
(but those could be improved by optimizing a detector sensitivity).  
%%% Because of such limitation of the gravitational wave observation, 
%%% it will be reasonable to parametrize nuclear EOS with the two segment 
%%% piecewise polytropic EOS for the case of triaxial compact stars.  

The stiffness of the EOS is the essential property affecting the maximum 
masses, and in our two segment piecewise polytropic EOS model that is 
parametrized by the $(\Gamma_1,\Gamma_2)$ of the outer and inner cores.  
Then, the possibilities are whether 
(i) the stiffness is approximately the same for inner and outer cores, 
$\Gamma_1 \approx\Gamma_2$, 
(ii) the inner core is stiffer than the outer core, $\Gamma_1<\Gamma_2$, or 
(iii) the inner core is softer than the outer core, $\Gamma_1>\Gamma_2$, 
and the stiffness of the outer core $\Gamma_1$ may be compared with 
$\Gamma \sim 2.5-3$ where the relativistic triaxial solutions appear.  
The case (i) is the same as a simple (one segment) polytropic EOS: 
a difference between axisymmetric and triaxial maximum masss 
defined by $\Delta M_{\rm max}:=(\Maximax-\Mtrimax)/\Msphmax$ 
will depend systematically on the indices $\Gamma_i$.  
The maximum mass difference $\Delta M_{\rm max}$ for the case (i) 
will be larger than $\Delta M_{\rm max} \agt 20\%$ in a range 
$2.24 \alt\Gamma \alt 4$ (and supporsedly in $\Gamma\agt 4$ also).  
For the case (ii), $\Delta M_{\rm max}$ can not be smaller than 
the case (i) because the maximum mass of spherical and axisymmetric 
solutions, $\Msphmax$ and $\Maximax$, are not affected by the EOS of 
the outer core but mostly by the inner core \cite{Read:2008iy}, while 
the maximum mass of triaxial star $\Mtrimax$ becomes smaller for 
the softer EOS in the outer core.  Hence $\Delta M_{\rm max}$ 
will be the same or larger than 20\% for the case (ii).  

Therefore it seems legitimate to conclude that the maximum mass 
difference $\Delta M_{\rm max}$ will be less than 10\% for outer core's 
$\Gamma_1\approx 4$, and inner core's $\Gamma_2 \alt 2.5$, 
and as considering the systematic dependence of the maximum masses on 
the stiffness of EOS, $\Delta M_{\rm max}$ would be around 10\% or less 
for the other combination of $\Gamma_i$, such as $\Gamma_1\approx 3.5$ and 
$\Gamma_2 \alt 2$ for the outer and inner cores, respectively.  
As an example, apart from the results presented in the previous section, 
we have also calculated a case with $(\Gamma_1,\Gamma_2)=(3.5,2.5)$, 
and found the mass difference to be $\Delta M_{\rm max}=15.4\%$.  
Clearly, 
such modifications in $\Gamma_i$ do not change the above statement, 
that the mass difference $\Delta M_{\rm max} \alt 10\%$ is a strong evidence 
for the softer inner core and stiffer outer core.  

The illustrative waveform in Fig.~\ref{fig:waveform} may be different 
from the actual wave form, 
because such compact rapidly rotating stars are also unstable to 
the Chandrasekhar-Friedman-Schutz (CFS) mechanism 
\cite{Chandrasekhar:1992pr,FriedmanSchutz,2013rrs..book.....F}, which sets in 
at a value of $T/|W|$ lower than that of the dominant viscosity-driven 
secular $\ell=m=2$ f-mode \cite{viscosityfmode}.  
Therefore, after $\Mtrimax$ is reached, we might still see the signals
of lower order gravitational f-modes ($m=2$-$4$) and/or r-modes 
\cite{1998MNRAS.299.1059A}.  Such modes are assumed to be suppressed 
in the above scenario because of a strong viscosity mechanism 
or turbulent magnetic flow.  If not, the modeling of the waveform 
becomes more challenging.  

Because of the recent successful detection of gravitational waves from 
a binary black hole merger, the detection of those from neutron stars 
looks very promising \cite{Abbott:2016blz}.  
Since the above signal from triaxial compact star resides 
roughly around 2000-3000Hz for a compactness 
$\compa\sim 0.2$-$0.3$ \cite{Saijo:2006um}, 
it will be necessary to improve the sensitivity in this bandwidth 
using narrow banding \cite{Hughes:2002ru}.  

\acknowledgments
This work was supported by 
JSPS Grant-in-Aid for Scientific Research(C) 15K05085, 25400262, 
26400267, and 26400274.  

\end{document}